\documentclass[a4paper, amsfonts, amssymb, amsmath, reprint, showkeys, nofootinbib, twoside]{revtex4-1}
\usepackage[english]{babel}
\usepackage[utf8]{inputenc}
\usepackage[colorinlistoftodos, color=green!40, prependcaption]{todonotes}
\usepackage{amsthm}
\usepackage{mathtools}
\usepackage{physics}
\usepackage{xcolor}
\usepackage{graphicx}
\usepackage[left=20mm,right=20mm,top=25mm,columnsep=15pt]{geometry} 
\usepackage{adjustbox}
\usepackage{placeins}
\usepackage[T1]{fontenc}
\usepackage{lipsum}
\usepackage{csquotes}
\newcommand{\jb}[1]{\textcolor{black}{#1}}

\usepackage[pdftex, pdftitle={Article}, pdfauthor={Author}]{hyperref} 
\begin{document}
\title{Two-step aging dynamics in enzymatic milk gels}

\author{Julien Bauland$^1$, Gouranga Manna$^2$, Thibaut Divoux$^1$, Thomas Gibaud$^1$}
    \email[Correspondence email address: ]{thomas.gibaud@ens-lyon.fr}
    \affiliation{$^1$ENSL, CNRS, Laboratoire de Physique, F-69342 Lyon, France}
    \affiliation{$^2$ESRF, The European Synchrotron, Grenoble 38043, France}

\date{\today} 

\begin{abstract}
Colloidal gels undergo a phenomenon known as physical aging, i.e., a continuous change of their physical properties with time after the gel point. 
To date, most of the research effort on aging in gels has been focused on suspensions of hard colloidal particles. In this letter, we tackle the case of soft colloidal ``micelles'' comprised of proteins, where gelation is induced by the addition of an enzyme. Using time-resolved mechanical spectroscopy, we monitor the viscoelastic properties of a suspension of colloidal micelles through the sol-gel transition and its subsequent aging. 
We show that the microscopic scenario underpinning the macroscopic aging dynamics comprises two sequential steps. First, the gel microstructure undergoes rapid coarsening, as observed by optical microscopy, followed by arrest. Second, aging occurs solely through a contact-driven mechanism, as evidenced by the square-root dependence of the yield stress with the elastic modulus measured at different ages of the gel. These results provide a comprehensive understanding of aging in enzymatic milk gels, which is crucial not only for a broad range of dairy products, but also for soft colloids in general.


\end{abstract}

\maketitle

Gels are two-component systems, composed of a small fraction of solid dispersed into a liquid phase and exhibiting solid-like properties. These soft solids find applications in countless fields, from ensuring the texture of food products~\cite{gibaud2012,Cao2020} to tissue engineering~\cite{Bertsch2023}, and environmental sciences~\cite{Baskaran2022}.

For particulate colloidal gels, the sol-gel transition results from the formation of a stress-bearing network due to attractive interactions between colloids. The network can be open and fractal or conversely compact and made of glassy strands, depending on the colloid volume fraction and the magnitude of attractive forces with respect to thermal energy~\cite{Trappe2000,Zaccarelli2007,gibaud2013,Johnson2019,Gibaud2020}. The structure and dynamics of the network bear crucial importance in determining the rheological properties of colloidal gels~\cite{Shih1990,Bouthier2022b,Bantawa2022,Cho2021}. Upon deformation, the reversibility of physical bonds and the multi-scale nature of the network result in many relaxation modes and rich viscoelastic spectra~\cite{Song2023}.

Beyond the gel point, colloidal gels show time-dependent properties referred to as ``aging'', where the gel elastic modulus increases as a function of time~\cite{Manley2005a,Ovarlez2007,Joshi2014,Gordon2017}. Aging has been attributed to two different microscopic origins. On the one hand, it may arise from structural rearrangements,  driven by either local reorganizations of the constituents due to thermal agitation~\cite{Cipelletti2000} or by larger scale or cooperative rearrangements resulting from internal stress relaxation~\cite{Bouzid2017,Jain,Begam2021}. On the other hand, a change in colloid interactions with time can lead to aging without any accompanying structural changes~\cite{Manley2005a,Bonacci2020,Bonacci2022}.

An example of particulate colloidal gel can be encountered in the field of food science, namely enzymatic milk gels.
Milk gelation is initiated through enzymatic destabilization of casein ``micelles'', which are natural colloids composed of proteins and salts~\cite{Holt2021}. Enzymatic milk gels display pronounced aging, which is a key feature during the early stages of cheese manufacturing~\cite{VanVliet1991,Fagan2017}. Despite its industrial significance  and numerous studies reporting on milk gel rheological properties, the aging mechanisms and their impact on the viscoelastic properties of enzymatic milk gels remain not fully understood~\cite{Mellema2002,Horne2017,Bauland2022a}. 

In this letter, we address the aging mechanisms of enzymatic milk gels using a combination of time-resolved mechanical spectroscopy and structural characterizations. Viscoelastic spectrum measured across the sol-gel transition allows us to build a master curve, which shows that enzymatic milk gels, tested at various volume fractions, obey a time-connectivity superposition principle akin to the time-cure superposition principle originally identified in polymer gels \cite{Winter1986,Martin:1988,Adolf1990}. Such a continuous evolution in the macroscopic response of enzymatic milk gels is related to extensive rearrangements of the gel microstructure, which lead to an increase of the fractal dimension of the gel network, as confirmed by X-ray scattering. We then identify a second aging regime, where the gel elasticity increases solely through a contact-driven
mechanism. This behavior is in stark contrast with gels made of hard colloids in which the aging scenario does not involve structural rearrangements of the network formed at the gel point~\cite{Keshavarz2021}. 



\begin{figure*}
    \includegraphics[scale=0.55, clip=true, trim=15mm 0mm 0mm 0mm]{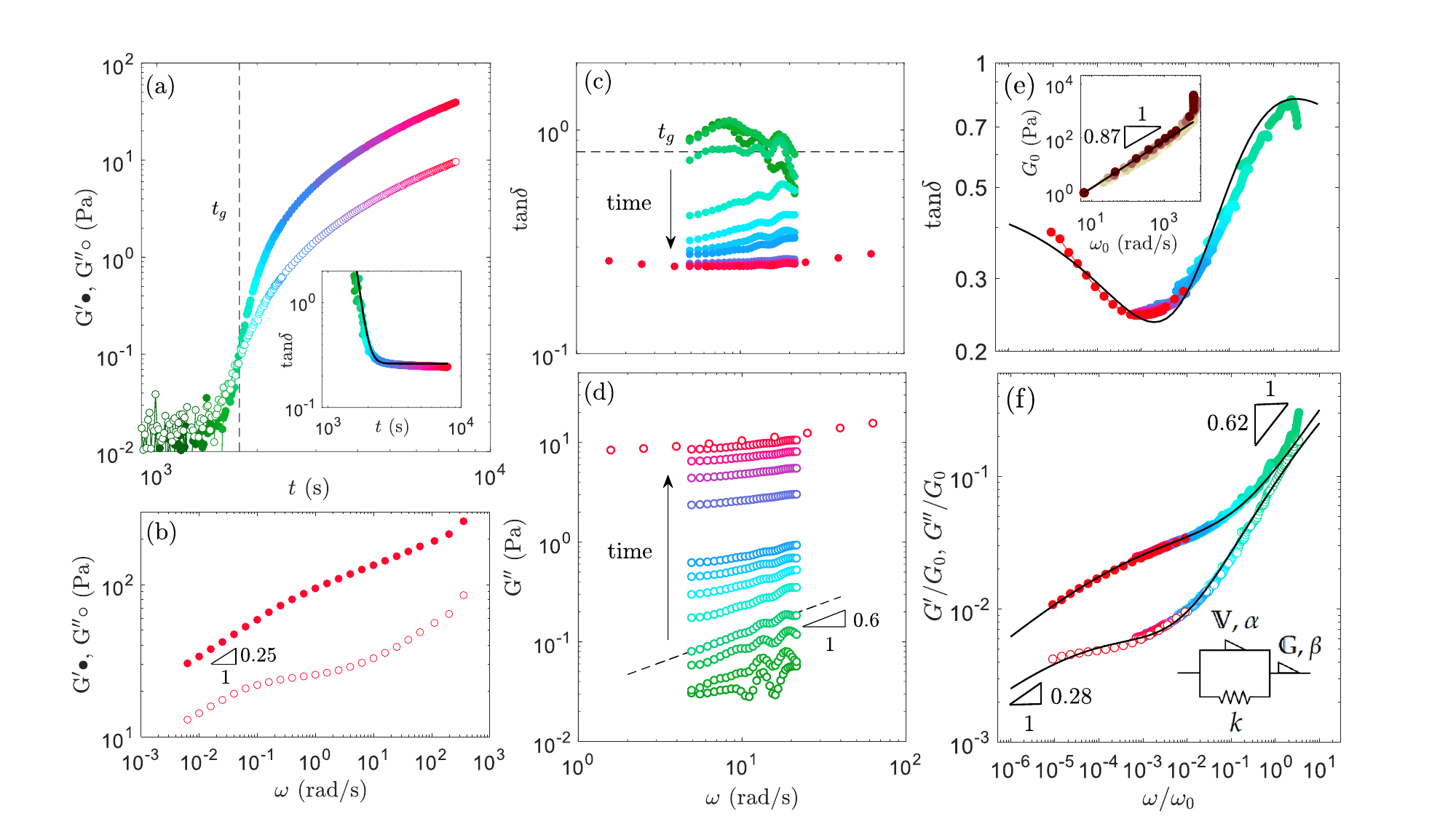}
    \centering
    \caption{Gelation from a rheology perspective. (a) Temporal evolution of elastic $G'$ (full) and viscous $G''$ (empty) moduli measured at angular frequency of  $\omega=0.628$ rad/s and a strain amplitude of $\gamma = 0.01$ after coagulant addition to the milk suspension at $\phi=10$~\%. The gel time $t_g = 1770~\rm s$ is indicated by the black dotted line. Inset: loss tangent $\tan \delta$ vs. time. The evolution of the loss tangent is fitted by an exponential decay (black line: $\tan \delta (t) \propto \exp[-(t-t_g)/\tau]$, with $\tau$ a characteristic time equals to 150 s for the 10 \% suspension. (b) Viscoelastic spectra of mature gel determined at $t=2.10^4$~s as a function of $\omega$. (c)-(d) Viscoelastic spectrum obtained from chirp measurements: (c) $\tan \delta$ and (d) $G^{\prime\prime}$ vs. $\omega$. Colors code for the aging time as depicted in (a). (e)-(f) Evolution of the loss tangent and rescaled moduli as a function of the rescaled angular frequency $\omega / \omega_0$. The black curves correspond to the best fit of the data with a fractional Poynting-Thomson model. Inset in (e) displays the scaling factors $G_0$ vs.~$\omega_0$ for different volume fractions of colloids: from light yellow to dark red for $\phi=$ 5, 7.5, 10, 15, and 20 \%.}
    \label{fig:chirp}
\end{figure*}


Reconstituted milk was obtained by dispersing a low-heat skim milk powder (Ingredia, Arras, France) in water to a mass fraction of 10 \% (w/w). The volume fraction $\phi$ of casein micelles was calculated from the casein concentration and was equal to approximately 10 \% (v/v). Subsequently, volume fractions ranging from 5 to 20 \% were prepared from the stock suspension (see SI for more details). Gelation was initiated at 30$^{\circ}$C by adding an enzymatic coagulant (Chr Hansen, Hoersholm, Denmark) to a final concentration of $0.05~\rm imcu.mL^{-1}$, thereby setting the time $t=0$~s (see SM). While still liquid, the suspension was introduced into a double-wall concentric cylinder geometry connected to a strain-controlled rheometer (ARES-G2, TA instrument) to perform rheological measurements through the sol-gel transition. In practice, linear viscoelastic spectra were measured periodically using multifrequency chirp (OWCh) signals to monitor the rapid temporal evolution of the elastic and viscous moduli (see SM)~\cite{Geri2018,Keshavarz2021}. In complement to rheological measurements, the evolution of the sample microstructure was determined by ultra-small-angle X-ray scattering (USAXS) (see SM), conducted at the European Synchrotron Radiation Facility (ESRF, ID02 beamline, France), and confocal microscopy imaging (see SM).

We first investigate the gelation kinetics of the casein micelles suspension with $\phi = 10$ \%. In Fig.~\ref{fig:chirp}(a), the elastic and viscous moduli $G^{\prime}(t)$ and $G^{\prime\prime}(t)$ quickly increase after a latency period of about 1000~s, which corresponds to the enzymatic destabilization and aggregation of casein micelles~\cite{Dekruif1998}. Concurrently, the loss tangent quickly decreases from $\tan \delta > 1$ to about 0.3, following an exponential-like decay $\tan \delta (t) =$ exp$[(t_g-t)/\tau] + k$, with $k$ the plateau value of $\tan \delta$(t) at long time and $\tau$, a characteristic time equal to $150~\rm s$ [inset in Fig.~\ref{fig:chirp}(a)]. 

From the OWCh protocol, viscoelastic spectra obtained throughout the sol-gel transition are displayed in Fig.~\ref{fig:chirp}(c). The gel point, defined as the time $t_g$ at which $\tan \delta$ is frequency independent \cite{Winter1986,Winter1997}, is found to be $t_g = 1770~\rm s$. At this critical point $G^{\prime}(\omega) \sim G^{\prime\prime}(\omega) \sim \omega^{\beta}$, with $\beta=0.62$ [Fig.~\ref{fig:chirp}(d) and Fig. 5(c) in SM]. This power-law response is linked to the self-similar nature of the percolated network at the gel point, which can be compactly described by its fractal dimension $d_f$. Beyond the gel point, the viscoelastic spectrum progressively evolves until it coincides with the spectrum of the mature gel obtained from a simple frequency sweep (initiated at $t \sim 2.10^4~\rm s$) displayed in Fig.~\ref{fig:chirp}(b).

Inspired by earlier reports on gels viscoelastic properties near the gel point~\cite{Larsen:2008,Chen:2010,Keshavarz2021,Costanzo2020,Morlet-Decarnin,Bantawa2022}, a single master curve can be constructed from the continuous evolution of $\tan \delta(\omega)$ by horizontally shifting the curves using a time-dependent scaling factor $\omega_0(t)$ [Fig.~\ref{fig:chirp}(e)]. In a similar way, $G^{\prime}(\omega / \omega_0)$ and $G^{\prime\prime}(\omega / \omega_0)$ can be rescaled by vertically shifting the curves using a second scaling factor $G_0(t)$ [Fig.~\ref{fig:chirp}(f)]. The scaling of the viscoelastic spectra acquired at different aging times points to a time-connectivity superposition principle, as reported for polymer \cite{Adolf1990, Adibnia2016} and particle gels \cite{Keshavarz2021,Morlet-Decarnin}. The time-connectivity principle highlights the self-similar evolution of the gel viscoelastic properties and suggests that changes taking place during aging essentially result in a change of scale.

The master curves $G^{\prime} / G_0$ and $G^{\prime\prime} / G_0$ vs. $(\omega /\omega_0)$ span over 6 orders of magnitude and display power-laws both in the high and low-frequency limits. Power-law responses are efficiently captured by fractional models and the introduction of a spring-pot element. Its constitutive equation follows $\sigma = \mathbb{V}d^{\alpha}\gamma/dt^{\alpha}$, with $\alpha$ a dimensionless exponents, that interpolates between a spring-like ($\alpha = 0$) and a dashpot-like ($\alpha = 1$) response. Moreover, $\mathbb{V}$ is a ``quasi-property'' with dimension Pa.s$^{\alpha}$ \cite{Koeller:1984,Bonfanti2020,Jaishankar2013}. Here, the full viscoelastic spectra can be captured by a Poynting-Thomson model~\cite{Bonfanti2020} composed of a spring-pot ($\mathbb{G}$, $\beta$) in series with a fractional Kelvin-Voigt element ($\mathbb{V}$, $\alpha$, and $k$) [see sketch as an inset in Fig.~\ref{fig:chirp}(f)]. By varying the volume fraction of casein micelles, we found that this evolution of the viscoelastic properties is robust for $5\leq \phi \leq 20$ \% and that the exponents $\beta = 0.62 \pm 0.04$ and $\alpha = 0.28 \pm 0.01$ are independent of $\phi$ [see Fig. 7(b) in SM].  

\begin{figure}
    \includegraphics[scale=0.48, clip=true, trim=0mm 0mm 0mm 0mm]{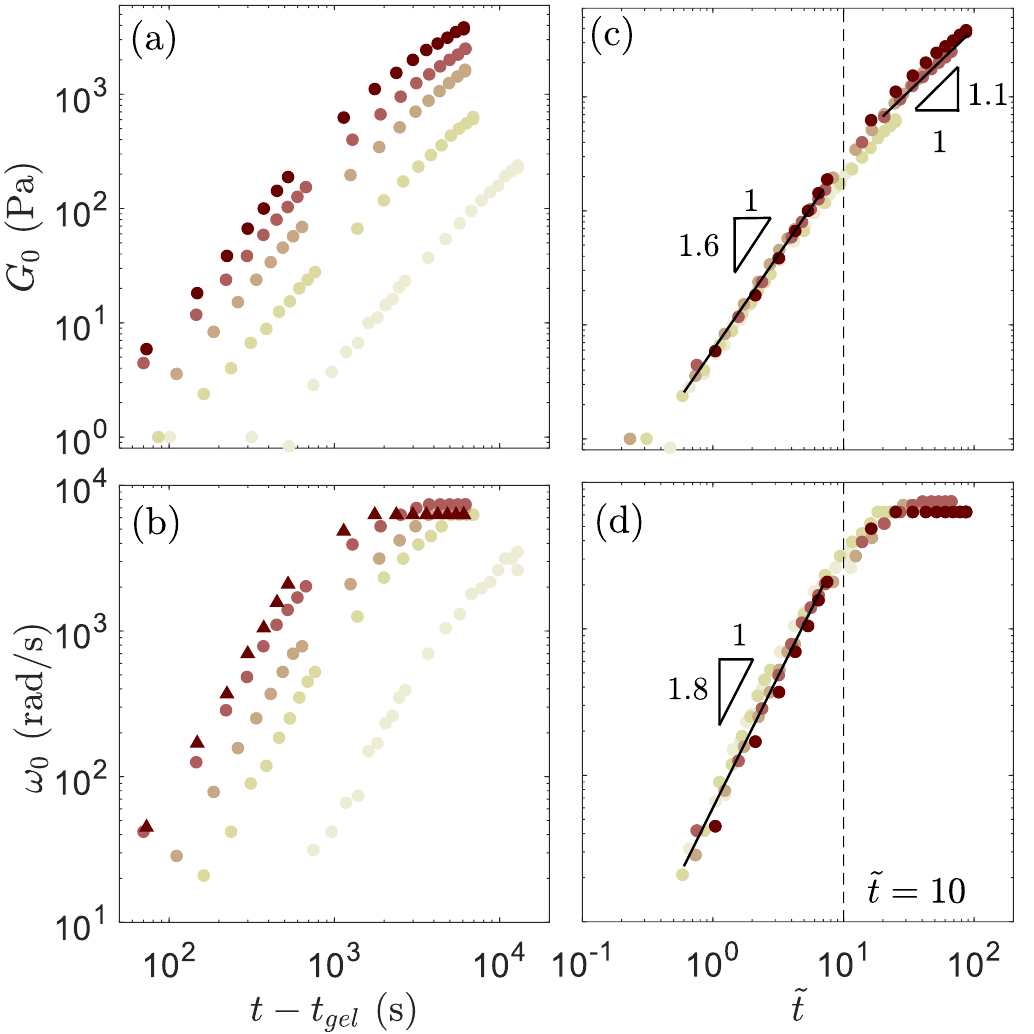}
    \centering
    \caption{Scaling factors vs. time at various volume fractions. (a)-(b) Scaling factors $G_0$ and $\omega_0$ as function of the elapsed time since the gel point $t-t_{g}$. Colors code for the volume fraction of colloids: from light yellow to dark red $\phi=$ 5, 7.5, 10, 15 and 20 \%. (c)-(d) $G_0$ and $\omega_0$ displayed as function of the reduced time $\tilde{t} = (t-t_{g}) / \tau$, with $\tau$ the characteristic time extracted from the exponential decay of $\tan \delta(t)$.}
    \label{fig:rescale}
\end{figure}

The time-dependence of the complete viscoelastic properties is contained in the evolution of the scaling factors $G_0(t)$ and $\omega_0(t)$,  the characteristic elasticity and representative time-scale of the material, respectively. They are displayed as a function of the time elapsed since the gel point, i.e., $t-t_{g}$ in Fig.~\ref{fig:rescale}(a) and \ref{fig:rescale}(b) for various volume fractions of casein micelles. 
All curves $G_0(t-t_{g})$ and $\omega_0(t-t_{g})$ collapse onto master curves [see Figs.~\ref{fig:rescale}(c) and \ref{fig:rescale}(d)], when plotted as a function of a dimensionless time $\tilde{t} = (t-t_{g})/\tau$, with $\tau$ the characteristic time extracted from the exponential decay of $\tan \delta$ with time [see inset in Fig.~\ref{fig:chirp}(a)]. First, such scaling indicates that, irrespective of the colloid volume fraction, the evolution of the viscoelastic properties of enzymatic milk gels during aging follows the same pathway. As $\tau$ decreases with the volume fraction according to $\tau \propto \phi^{-0.4}$ [see Fig. 7(a) in SM], only the speed of the aging process increases with the volume fraction. Second, the evolution of $G_0(\tilde{t})$ and $\omega_0(\tilde{t})$ shows two distinct dynamics, separated by a common critical time $\tilde{t}_c \approx 10$. For $\tilde{t} < \tilde{t}_c $, both $G_0$ and $\omega_0$ increase as power laws of $\tilde{t}$ with exponents of $1.6$ and $1.8$, respectively. In contrast, for $\tilde{t} \geq \tilde{t}_c $, $\omega_0$ is constant and $G_0$ increases almost linearly with $\tilde{t}$. 

\begin{figure}
    \includegraphics[scale=0.68, clip=true, trim=0mm 0mm 0mm 0mm]{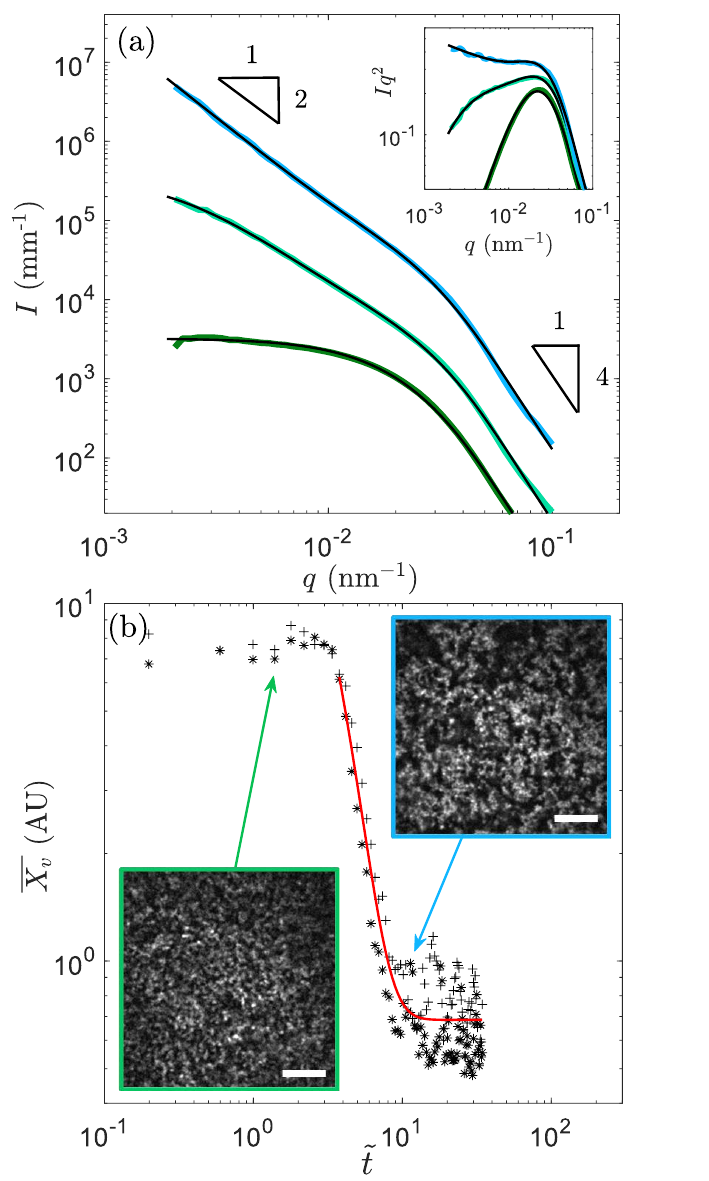}
    \centering
    \caption{Evolution of the microstructure through the sol-gel transition and subsequent aging. (a) USAXS intensity $I$ vs.~the wave vector $q$ at selected aging times: $\tilde{t}$ = -6, 0, and 10 from bottom to top. For clarity, horizontal shift factors equal to 10 and 100 are applied for $\tilde{t}$ = 0 and 10, respectively. Inset: Kratky plot $I.q^2$ vs.~$q$. (b) Magnitude of displacement vectors $\overline{X}_v$ computed by cross-correlation of confocal images recorded during the sample aging. $\overline{X}_v$ is taken as an estimate of structural change between two successive images. Red curve represents the best exponential decay fit of the data. Inset: confocal images at $\tilde{t}$ = 0 and 10. The scale bar represents 10 \textmu m.
    }
    \label{fig:struct}
\end{figure}

To gain insights into the microscopic mechanisms underlying this two-step aging process separated by $\tilde{t}_c$, we probe the gel structure at $\phi = 10$ \% using USAXS and confocal microscopy. Fig.~\ref{fig:struct}(a) displays the scattered intensity $I(q)$ plotted against the wave vector $q$, up to $\tilde{t} \approx \tilde{t}_c$. At $\tilde{t} = -6$ (i.e., immediately after coagulant addition), the scattering intensity of the stable casein micelles suspension shows a Guinier plateau at low $q$ attributed to the spherical form factor of the colloids \cite{Bouchoux2010}. At high $q$, the intensity decreases according the Porod's law, expressed as $I(q) \sim q^{-4}$, and is attributed to the scattering of the colloids interface. After coagulant addition, while time increases, the enzymatic reaction proceeds and casein micelles structure into fractal clusters. As a result, the scattering intensity at low $q$ increases due to the volume increase of the scatterers. For $q<0.02$~nm$^{-1}$, the scattered intensity shows a power-law scaling $I(q) \sim q^{d_f}$ that can be captured by a mass fractal structure factor~\cite{Teixeira1988}. At the gel point [i.e., $\tilde{t} = 0$; light green curve in Fig.~\ref{fig:struct}(a)], the fractal dimension of the network formed by the percolated micelles is $d_f = 2$ and increases to $d_f = 2.2$ at $\tilde{t} = 10$ [blue curve in Fig.~\ref{fig:struct}(a)]. This change of fractal dimension during aging, as reflected by a change in $I(q)$, is better visualized when data are displayed using Kratky representation [see inset in Fig.~\ref{fig:struct}(a)]. The increase of $d_f$ over time indicates that the percolated network of casein micelles spontaneously coarsens during aging. \jb{This scenario closely resembles the restructuring of polystyrene latex particle gels probed by light scattering \cite{Cipelletti2000}.} 

To quantify the dynamics of the structural changes occurring at length scales larger than that accessible by X-ray scattering, we monitor the evolution of enzymatic milk gels by confocal microscopy imaging (see movie 1 in SM).
Casein micelles that appear in white, quickly structure into freely diffusing clusters. These clusters percolate and their dynamics become arrested at a critical time beyond which changes in the microstructure appear local. Using cross-correlation analysis, we track the displacement of the gray-scale structures over successive images using a strategy developed for particle imaging velocimetry \cite{Thielicke2021}, and previously employed to study syneresis in colloidal gels \cite{Wu2080}. The magnitude of the displacement vector $\overline{X}_v$, displayed as a function of $\tilde{t}$ in Fig.~\ref{fig:struct}(b), is taken as a metric of structural changes between two successive images (see movie 2 in SM). When the suspension is liquid, $\overline{X}_v$ is constant and its value is attributed to random correlations between images.
In the vicinity of the gel point ($\tilde{t} \approx 1$), correlations between pair images increase due to the structuring of casein micelles into clusters, which give rise to a small overshoot in $\overline{X}_v$ as clusters still retain a high degree of mobility. For $\tilde{t} > 5$, $\overline{X}_v$ drops abruptly, as the structure displacements slow down to eventually stop when the gel is fully formed. Finally, for $\tilde{t} > \tilde{t}_c=10$, $\overline{X}_v$ is constant and the method does not detect any additional changes in the gel microstructure for length scales superior to $0.1$ \textmu m, which is the spatial resolution of the correlation analysis. The setting of the gel microstructure at $\tilde{t_c} = 10$ is fully consistent with the change in the dynamics of $G_0(\tilde{t})$ and $\omega_0(\tilde{t})$ depicted in Figs.~\ref{fig:rescale}(c) and \ref{fig:rescale}(d). Comparison of images taken at $\tilde{t} =$ 1 and 10 [insets in Fig.~\ref{fig:struct}(b)] shows the evolution of finely distributed clusters at $\tilde{t} =$ 0 to a coarse network, consistent with the increase of the fractal dimension measured in USAXS. 

Building upon these observation, we can propose a microscopic scenario accounting for the two-step aging dynamics of enzymatic milk gels. The first step is dominated by a coarsening of the gel network that involves large-scale rearrangements of the microstructure. It leads to a fast increase of the gel elasticity $G_0$ up to the time $\tilde{t}_c =10$, which corresponds to an arrest transition, and a ``freezing'' of the microstructure that ceases to rearrange. For $\tilde{t} >\tilde t_c$, we observe that, despite the lack of microscale rearrangements, the gel elastic properties continue to reinforce (i.e., $G_0 \sim \tilde{t}^{1.1}$). This observation suggests that mechanical aging is now driven by some local aging process. Here, we hypothesize that the macroscopic mechanical aging is due to contact-driven aging between neighboring casein micelles, as postulated in previous work on enzymatic milk gels~\cite{Mellema2002,Bauland2022a}.

\begin{figure}
    \includegraphics[scale=0.55, clip=true, trim=2mm 22mm 2mm 32mm]{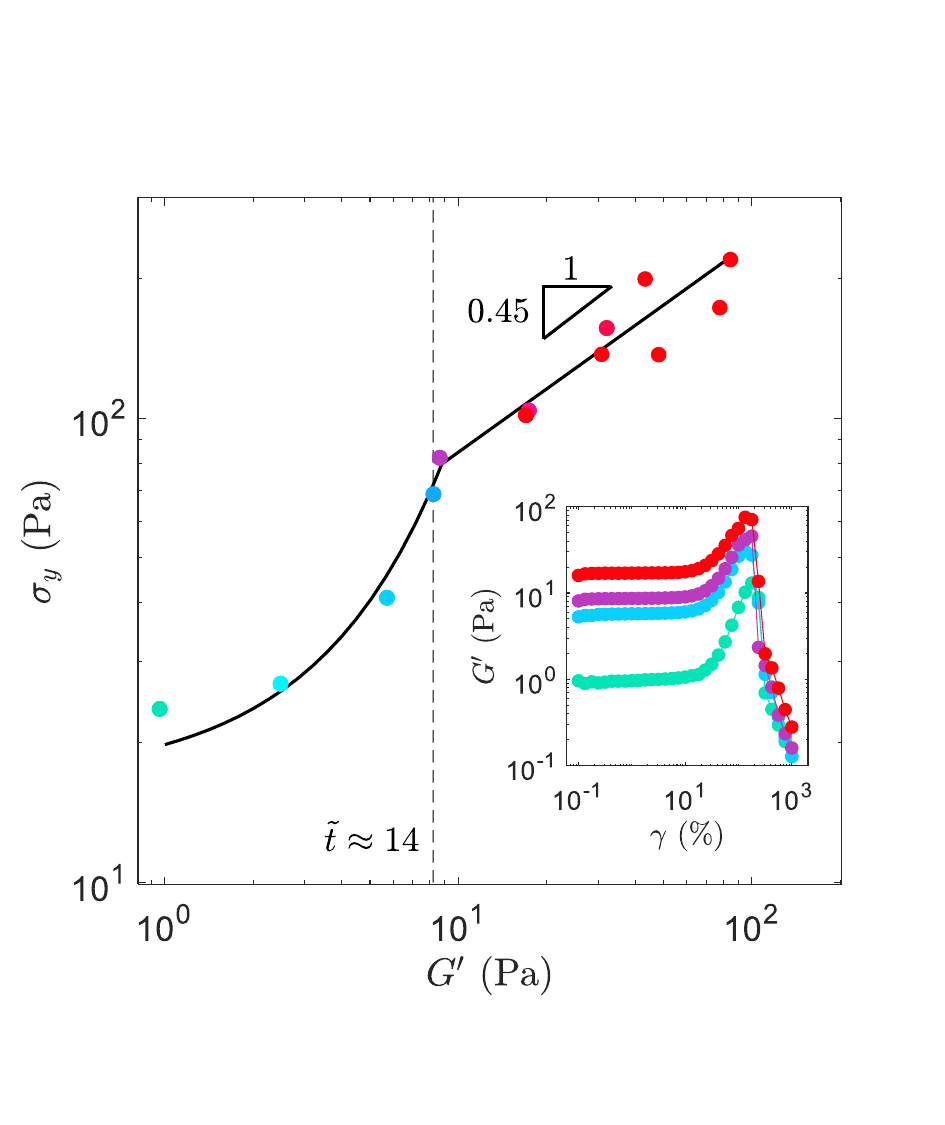}
    \centering
    \caption{Yield stress $\sigma_y$ vs. elastic modulus $G^{\prime}$ measured at different ages. Colors code for the aging time as in Fig.~\ref{fig:chirp}. Solid lines represent the best exponential and power-law fits of the data before and after $\tilde{t} \approx 14$, respectively. Inset displays strain sweep tests used to determine $\sigma_y$}
    \label{fig:sweep}
\end{figure}

In fact, contact-driven aging dynamics has been previously reported in gels of solid colloidal particles~\cite{Manley2005a}, and more recently in dense suspensions of silica particles~\cite{Bonacci2020}, where a constitutive relation between the shear modulus and the yield stress was established by determining the sample mechanical properties at various ages~\cite{Bonacci2022}. In case of contact-aging, the yield stress of the sample is expected to scale with the square root of the shear modulus, i.e., $\sigma_y \propto \sqrt{G^{\prime}}$~\cite{Bonacci2022}. To test whether such a relation is valid in gels of casein micelles, we conducted stress sweep tests at various aging times (see inset in Fig.~\ref{fig:sweep}). Each experiments performed on a fresh gel yields an elastic modulus (at low strain values) and a yield stress $\sigma_y$ determined by the abrupt drop in $G'$. We report $\sigma_y$ vs.~$G'$ in Fig.~\ref{fig:sweep}, where the gel age is indicated by the same color code as in Figs.~\ref{fig:chirp} and ~\ref{fig:struct}. At short aging times ($\tilde{t} < \tilde t_c$), $\sigma_y$ increases exponentially with $G'$, whereas for $\tilde{t} \geq \tilde t_c$, we find that $\sigma_y \propto G^{\prime \lambda}$ with $\lambda = 0.45 \pm 0.05$, in very good agreement with the square root prediction expected for contact-driven aging~\cite{Bonacci2022}. This result confirms that once the microstructure structure of enzymatic milk gel is frozen, the strengthening of the macroscopic mechanical properties of the gel proceeds through contact-driven aging.


Let us now summarize and discuss the key findings of this Letter. Using a multiscale approach, we have shown that the aging of enzymatic milk gels displays two distinct dynamics, irrespective of the volume fraction of casein micelles. The first regime is dominated by structural rearrangements so that the network formed at the gel point coarsen, without losing its fractal nature. By comparing rheological and scattering data, it is possible to relate the fractal dimension of network and its power-law rheology, characterized by a single exponent $\Delta$, using Muthukumar's relation \cite{Muthukumar1989}. Derived in the context of polymer gels, the relation follows $\Delta = 3(d+2-2d_f) / 2(d+2-d_f)$, with $d$ the euclidean dimension. Testing this relation with the exponents reported above for the critical gel ($\Delta=\beta = 0.62$) and the mature gel ($\Delta=\alpha = 0.28$) leads, respectively, to $d_f \approx 1.9$ at the gel point, and $d_f \approx 2.2$ for the mature gel, in good agreement with the values extracted from USAXS [see Fig.~\ref{fig:struct}(a)]. It demonstrates that structural rearrangements directly account for the evolution of the gel viscoelastic properties. This scenario contrasts with previous report on aluminosilicate and silica particle gels, where the structure of the network formed at the critical gel point was preserved during aging~\cite{Keshavarz2021}. The preservation of the percolated network structure throughout aging is consistent with the framework developed by Adolf and Martins~\cite{Adolf1990}, where the time-connectivity superposition principle relies on the self-similarity of the network at the gel point, and the fact that the effect of aging essentially result in a change of scale. In the present case, despite the rearrangement of the structure formed at gel point, the time-connectivity principle still applies [Fig.\ref{fig:chirp}(b) and \ref{fig:chirp}(f)], highlighting its robustness, yet questioning its physical interpretation.

Following the coarsening of the network, the mesoscopic structure was found to “freeze” at a critical time $\tilde{t}_c = 10$. From this critical time, aging proceeds by contact aging solely, as evidenced by the scaling of the yield stress with the square root of the elastic modulus. In this second regime, the elasticity of the gel increases linearly with $\tilde{t}$ [Fig.~\ref{fig:rescale}(c)], which is faster than the power-law increase $G^{\prime} \sim t^{0.4}$ reported for silica particle gels~\cite{Manley2005a}. 

It can be argued that contact-aging between casein micelles probably starts immediately after aggregation, but does not preclude structural rearrangements. In fact, contact-aging has been postulated as the motor of physical aging in enzymatic milk gels, as suggested by the sintering of casein micelles observed by electron microscopy~\cite{Mellema2002, Bauland2022a}. Casein micelles are characterized by a Young modulus of about $10^5$ Pa~\cite{Bauland2022}, and are significantly softer than colloidal particles usually employed to study colloidal gels ({e.g.}, $10^{10}$ Pa for silica particles). More deformable particles were found to favor the local compaction of colloidal gels, due to a lower contribution of the bending mode to the stress relaxation~\cite{Thiel2020a,Wu2080}. Accordingly, local rearrangements of casein micelles due to contact aging may induce internal stresses on the percolated network and large scales rearrangements during the early stages of aging (i.e., when $\tilde{t} < \tilde{t}_c$). The existence of internal stresses in enzymatic milk gels is supported by their strong syneresis properties, as internal stresses are thought to be responsible for the pressure induced on the solvent during its expulsion~\cite{VanVliet1991,leocmach2015}. Our results strongly recalls the aging dynamics studied using dynamic light scattering in gels of soft latex colloids~\cite{Cipelletti2000}, where eventually sintering between the particle was also reported, supporting the generality of our two-step aging scenario for a broad range of soft colloids.

This study offers a comprehensive understanding of the aging scenario in enzymatic milk gels and provides a benchmark set of data for future theoretical studies regarding aging in gels made of soft attractive colloidal particles~\cite{Vlassopoulos2014}.


The authors thank Ingredia (Arras, France) for providing the low heat skim milk powder and Chr. Hansen (Hoersholm, Denmark) for providing the enzymatic coagulant. This work was supported by the LABEX iMUST of the University of Lyon (ANR-10-LABX-0064), created within the ``Plan France 2030" set up by the French government and managed by the French National Research Agency (ANR). We acknowledge the contribution of Elodie Chatre and the SFR Biosciences (UAR3444/CNRS, US8/Inserm, ENS de Lyon, UCBL) facilities for the confocal microscopy data. Special thanks to ESRF and beamline ID02 (SC-5455) for granting beamtime.

\section*{Supplementary Materials}

\subsection*{Sample preparation}

Reconstituted milk was obtained by dispersing a low-heat skim milk powder (Ingredia, Arras, France) in water to a mass fraction of 10 \% (w/w). Sodium azide (NaN$_3$) was added as bio-preservative to a final concentration of 0.2 g.L$^{-1}$. As commonly performed in dairy processing, 2 mM of calcium chloride (CaCl$_2$) were also added to enhance gelation~\cite{Bauland2022a}. The suspensions was agitated under magnetic stirring at room temperature during 12~h. The volume fraction $\phi$ of casein micelles was calculated from the casein concentration, taken as $\sim30$ g.L$^{-1}$ according~\cite{Bauland2022a} and the voluminosity of casein micelles, taken as 3 mL.g$^{-1}$ of dry casein~\cite{Walstra1979}. 

Milk samples with volume fractions of casein micelles within the range $5 \leq \phi \leq 20$~\% (v/v) were prepared from the 10~\% stock suspension by mean of ultra-centrifugation. Casein micelles were concentrated by centrifuging the stock suspension at 70,000 $g$ during 1~h at 20$^{\circ}$C. To produce suspensions with volume fractions higher than 10~\%, the casein micelle pellets were re-dispersed in the stock suspension and left overnight under magnetic stirring. To produce suspensions with a volume fractions lower than 10 \%, the stock suspension was diluted with its own solvent, i.e., the supernatant obtained by ultra-centrifugation.

Gelation was initiated at 30$^{\circ}$C by adding an enzymatic coagulant (Chr Hansen, Hoersholm, Denmark) to a final concentration of $0.05~\rm imcu.mL^{-1}$. The enzymatic coagulant is mainly composed of \emph{chymosin}, a proteolytic enzyme responsible for the specific hydrolysis of the $\kappa$-casein at the surface of the casien micelles. Such hydrolysis hinders the repulsion between casein micelles~\cite{Horne2017} and destabilizes the dispersion.

\subsection*{OWchirp protocol} 

\begin{figure*}
    \includegraphics[scale=0.44, clip=true, trim=0mm 0mm 0mm 0mm]{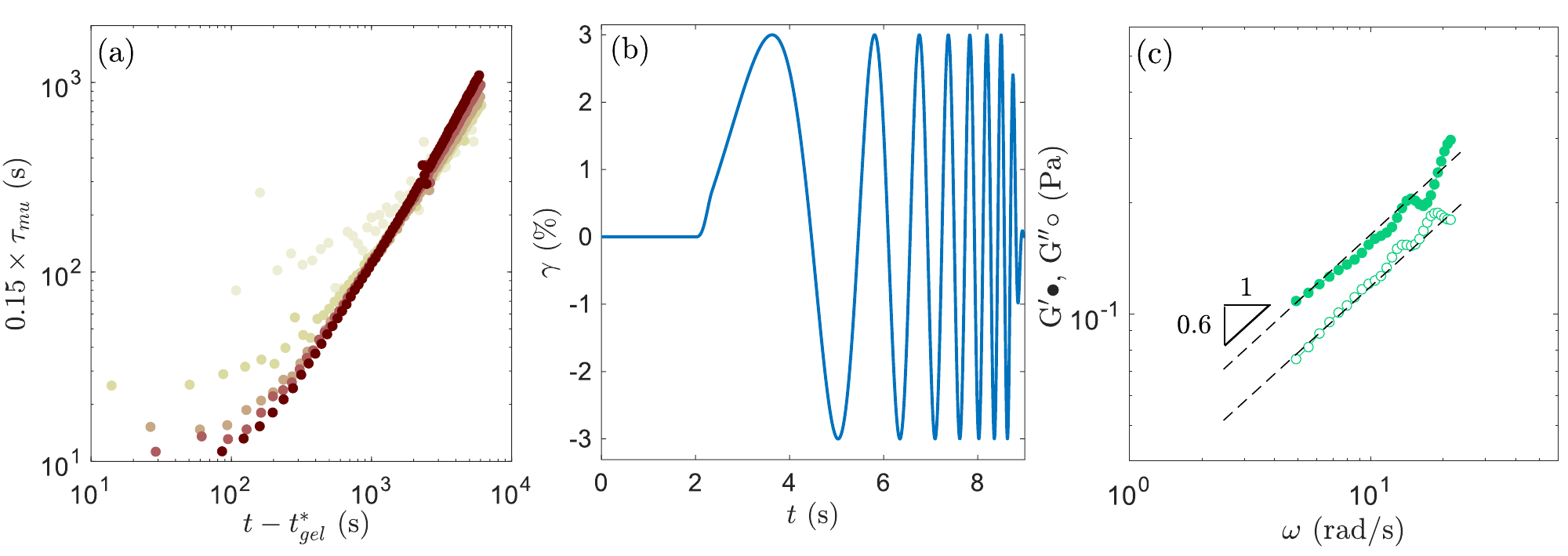}
    \centering
    \caption{(a) Mutation time $\tau_{mu}$ of the enzymatic milk gels calculated from the aging kinetics measured by standard time sweep test: $\tau_{mu} = ( dG^{\prime}/dt)^{-1}$. The fraction $0.15\times \tau_{mu}$ is displayed vs. the time elapsed since the gel point $t_g$, which directly indicates the maximum duration of the test that can be applied at this time of the aging kinetics. (b) OWChirp sequence (strain $\gamma$ vs. time) designed using the graphical user interface developed in \cite{Geri2018}. (c) Viscoelastic spectra of the 10 $\%$ (v/v) casein micelle suspension measured at the critical gel point with the sequence displayed in (b). Black dotted lines indicate the best power-law fit of the data, which follows $G^{\prime}(\omega) \sim G^{\prime\prime}(\omega) \sim \omega^{0.6}$.
    }
    \label{fig:OWChirp}
\end{figure*}

The temporal evolution of the viscoelastic properties of enzymatic milk gels were measured periodically using Optimally Windowed Chirp (OWCh) signals \cite{Geri2018}. At short aging times (i.e., $t \approx t_g$), the mutation time of the gel $\tau_{mu} = ( dG^{\prime}/dt)^{-1}$ is about 60 s [Fig.~\ref{fig:OWChirp}(a)]. To ensure that the test duration remains a small fraction of the mutation time, taken as $0.15 \times \tau_{mu}$ \cite{Geri2018}, a chirp sequence with a duration of 9 s [Fig.~\ref{fig:OWChirp}(b)] was chosen to collect reliable data on a frequency range comprised between 4 and $20~\rm rad/s$ [Fig.~\ref{fig:OWChirp}(c)]. Between chirp sequences, the gelation kinetics was measured by single wave oscillation rheometry with $\omega=0.628~\rm rad/s$ and $\gamma=0.01$.

\subsection*{Confocal microscopy} 

Fast imaging was performed on an inverted spinning disk confocal (Leica, Yokogawa) equipped with a thermoregulated chamber. Caseins were stained using Fast green FCF with a ratio casein/dye of $2.4 \times 10^3$ (w/w) following the protocol described in \cite{Bauland2023}.

Movie 1 displays confocal images of the casein micelles suspension ($\phi = 10$ \%) through the sol-gel transition and subsequent aging. For each image, the corresponding rescaled time $\tilde{t}$ is indicated and the scale bar represents 10 \textmu m. Casein micelles that appear in white, quickly structure into freely diffusing clusters. These clusters percolate into a network whose structure coarsen before becoming completely arrested at a critical time beyond which changes in the microstructure appear local. Movie 2 displays the same set of images with the displacement vectors (green arrows) calculated from cross-correlation analysis. The magnitude of the displacement vectors decreases as the gel microstructure becomes arrested.

\subsection*{Ultra small-angle X-ray scattering} 

The microstructural characteristics of the casein micelles suspension with $\phi = 10$ \% were probed using USAXS measurements conducted at the ID02 beamline within the European Synchrotron Radiation Facility (ESRF) in Grenoble, France~\cite{Narayanan2022}. The incident X-ray beam, with a wavelength close to 0.1~nm ($E=12.23$ keV), was collimated to dimensions of 50~$\mu$m vertically and 100~$\mu$m horizontally. Utilizing an Eiger2 4M pixel array detector, two-dimensional scattering patterns were acquired. The scattering intensity $I(q)$ was derived by subtracting the two-dimensional scattering profiles of the casein suspension and the aqueous background (taken as the supernatant of ultra-centrifugation). The resulting scattering intensity remained isotropic throughout this study and an azimuthal average was performed to obtain a one-dimensional $I(q)$. 

\begin{figure}
    \includegraphics[scale=0.55, clip=true, trim=0mm 0mm 0mm 0mm]{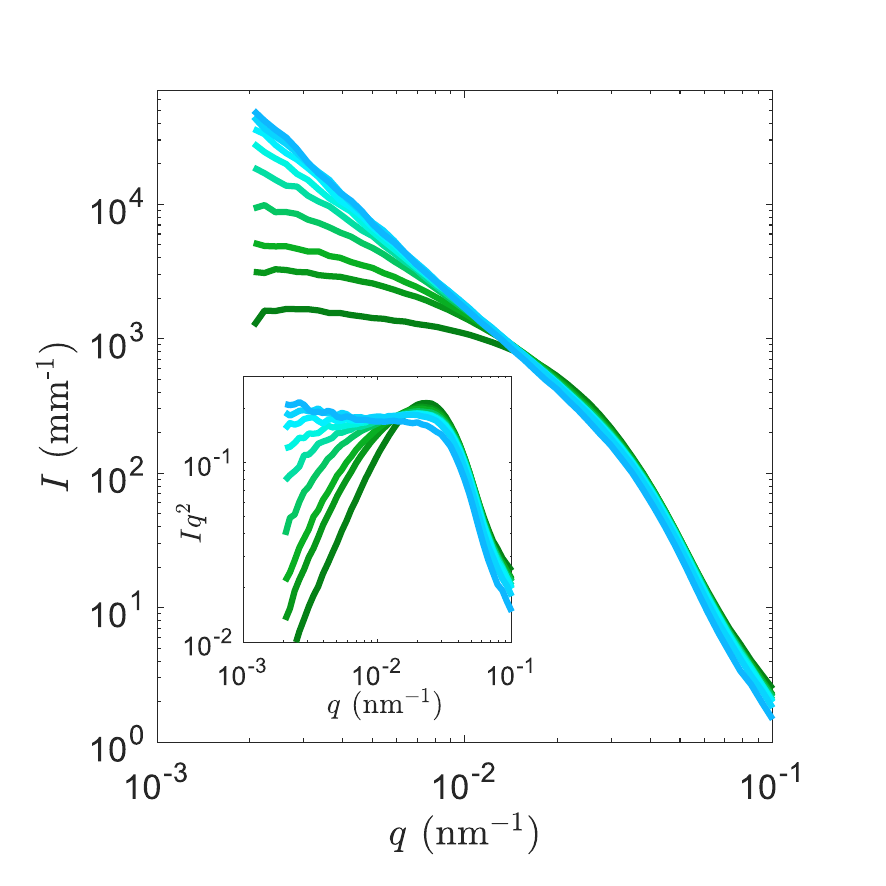}
    \centering
    \caption{1D scattering spectra $I(q)$ vs.~$q$ of the 10~\% casein micelle suspension acquired during gelation and aging. Colors code for the rescaled aging time $\tilde{t} = (t - t_g) / \tau$ defined in the main text. From bottom to top: $\tilde{t} =$ -7, -3, -2, -1, 0, 1, 2, 5, and 10. Inset displays the same data set in a Kratky plot, i.e., $I(q).q^2$ vs.~$q$.}
    \label{fig:SAXS}
\end{figure}

Fig.~\ref{fig:SAXS} displays the scattered intensity $I(q)$ vs.~$q$ for the 10 $\%$ casein micelle suspension at different aging times. The intensity spectrum $I(q)$ was fitted by the following model: $ I\left(q\right)= \phi \left(\Delta \rho\right)^2V^2 P\left(q\right) S\left(q\right)$, where $\phi$ is the volume fraction of casein micelles, $\Delta \rho$ is the difference of scattering length density between the casein micelles and the solvent, and $V$ the volume of casein micelles. $P(q)$ is the form factor of the casein micelles, modeled as polydisperse spheres with a Schultz distribution \cite{Aragon1976}. $S(q)$ is a mass fractal structure factor as given by Teixeira \cite{Teixeira1988}:

\begin{equation}
S(q) = 1 + \frac{d_f\Gamma(d_f-1)}{[1 + 1/(q\xi)^2]^{(d_f-1)/2}}\frac{\sin[(d_f-1)\tan^{-1}(q\xi)]}{(qR_0)^{d_f}}
\end{equation}

The scattering length density of the milk aqueous phase $\rho_s=9.84.10^{10}$ and of the casein micelles $\rho_c=12.7.10^{10}$ cm$^{-2}$ were taken from \cite{Bouchoux2010}. The volume fraction of casein micelles $\phi$ was a fitting parameter and was found to be $\phi \sim 0.09$ for the liquid suspension, consistent with the volume fraction expected from sample preparation. When the suspension structures into clusters, $\phi$ varies between 9 \% and 7 \%, resulting from the decrease of the scattering intensity of the form factor at $q \sim 3.10^{-3}~\rm nm^{-1}$.

\subsection*{Effects of volume fraction on gelation and aging kinetics} 

\begin{figure*}
    \includegraphics[scale=0.6, clip=true, trim=6mm 0mm 0mm 0mm]{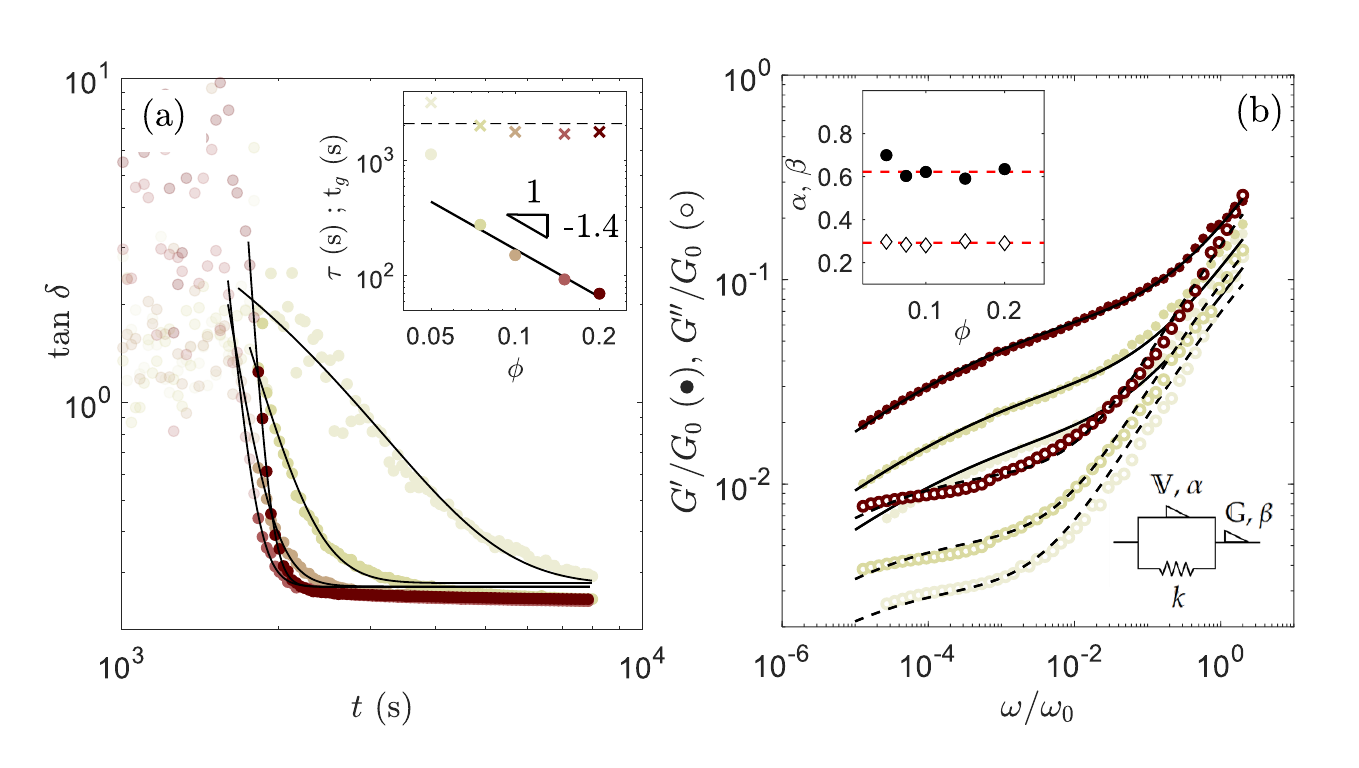}
    \centering
    \caption{(a) Temporal evolution of the loss tangent $\tan \delta$ after coagulant addition ($\omega=0.628~\rm rad/s$, $\gamma = 0.01$) for various volume fractions. From light yellow to dark red $\phi=$ 5, 7.5, 10, 15, and 20 \%. Black curves are the best fit of the data with an exponential-like decay $\tan \delta (t) =$ exp$[(t_g-t)/\tau] + k$, with $k$ the plateau value of $\tan \delta$(t) at long time and $\tau$ a characteristic time. Inset: characteristic time $\tau$ (full circles) and gel time $t_g$ (crosses) as function of the volume fraction. Black curve is the best power-law fit of $\tau$ vs.~$\phi$. (b) Evolution of the rescaled moduli as a function of the rescaled frequency $\omega / \omega_0$ for $\phi=$ 0.05, 0.075 and 0.2. Colored curves correspond to the best fit of the data with a fractional Poynting-Thomson model. Inset displays the fractional exponents $\beta$ and $\alpha$ as function of the volume fraction.}
    \label{fig:volfrac}
\end{figure*}

Fig.~\ref{fig:volfrac}(a) displays the loss factor $\tan \delta (t)$ during gelation and aging of casein micelle suspensions for $5 \leq \phi \leq 20$~\%, measured by small amplitude oscillations. The gel time remained almost independent of the volume fraction of casein micelles [$t_g \in [1700 - 3000]$ s in inset of Fig.~\ref{fig:volfrac}(a)]. As introduced in the manuscript, the temporal evolution of the loss tangent [Fig.~\ref{fig:volfrac}(a)] can be approximated by an exponential-like decay $\tan \delta(t) = \exp[(t_g-t)/\tau] + k$, with $k$ the value of $\tan \delta (t)$ at long time and $\tau$ a characteristic time. This characteristic time $\tau$ shows a power-law behavior with the volume fraction $\tau \sim \phi^{-1.4}$ [inset in Fig.~\ref{fig:volfrac}(a)].

Fig.~\ref{fig:volfrac}(b) displays the master curves obtained from the scaling of the viscoelastic spectra acquired at different aging times for $\phi=$ 0.05, 0.075, and 0.2. The master curves share the same features for all tested volume fractions. The fitting with the Poynting-Thomson fractional model yields $\phi$-independent fractional exponents $\beta = 0.62 \pm 0.04$ and $\alpha = 0.28 \pm 0.01$ [see inset in Fig.~\ref{fig:volfrac}(b)].

\bibliographystyle{apsrev4-1}
%

\end{document}